# Study of fusion-fission in inverse kinematics with a fragment separator

O. B. Tarasov,[1,*]  O. Delaune,[2,†]  F. Farget,[2]  D. J. Morrissey,[1,3]  A. M. Amthor,[4]  B. Bastin,[2] D. Bazin,[1]  B. Blank,[5]  L. Cacéres,[2]  A. Chbihi,[2]  B. Fernández-Dominguez,[6]  S. Grévy,[5] O. Kamalou,[2]  S. M. Lukyanov,[7]  W. Mittig,[1,8]  J. Pereira,[1]  L. Perrot,[9]  M.-G. Saint-Laurent,[2]  H. Savajols,[2]  B. M. Sherrill,[1,8]  C. Stodel,[2]  J. C. Thomas,[2]  A. C. Villari [10]

[1] *National Superconducting Cyclotron Laboratory, MSU, East Lansing, MI 48824, USA*
[2] *Grand Accelerateur National d'Ions Lourds, CEA/DSM-CNRS/IN2P3, F-14076 Caen, France*
[3] *Department of Chemistry, Michigan State University, East Lansing, MI 48824, USA*
[4] *Department of Physics, Bucknell University, Lewisburg, PA 17837, USA*
[5] *CENBG, UMR 5797 CNRS/IN2P3, Université Bordeaux 2, F-33175 Gradignan, France*
[6] *Universidade de Santiago de Compostela, E-15782 Santiago de Compostela, Spain*
[7] *FLNR, JINR, Dubna, Moscow region, 141980, Russian Federation*
[8] *Department of Physics and Astronomy, MSU, East Lansing, MI 48824, USA*
[9] *IPN Orsay, CNRS/IN2P3, F-91406 Orsay, France*
[10] *Facility for Rare Isotope Beams, MSU, East Lansing, MI 48824, USA*

The systematic study of fission fragment yields under different initial conditions provides a valuable experimental benchmark for fission models that aim to understand this complex decay channel and to predict reaction product yields. Inverse kinematics coupled to the use of a high-resolution spectrometer is shown to be a powerful tool to identify and measure the inclusive isotopic yields of fission fragments. In-flight fusion–fission was used to produce secondary beams of neutron-rich isotopes in the collision of a $^{238}$U beam at 24 MeV/u with $^{9}$Be and $^{12}$C targets at GANIL using the LISE3 fragment-separator. Unique $A,Z,q$ identification of fission products was attained with the d$E$-TKE-B$\rho$-ToF measurement technique. Mass, and atomic number distributions are reported for the two reactions that show the importance of different reaction mechanisms for these two targets.

## 1. Introduction

### 1.1. *Fusion–fission is a new mechanism to produce rare isotope beams*

Pioneering in-flight fission experiments at GSI intensively explored neutron-rich isotopes with $Z = 28 - 60$ [1]. Fission is and has been widely used to produce rare neutron-rich nuclei using different mechanisms to induce the fission process (abrasion-fission, Coulomb fission) combined with in-flight separation as well as spallation reactions with thick Uranium targets and ISOL techniques to produce neutron-rich isotopes of elements with $60 < Z < 70$. Production techniques for these nuclei using heavy targets in so-called normal kinematics suffers from difficulties with fragment extraction from the target and identification of the slow moving fragments. On the other hand, in-flight fusion–fission using reverse kinematics can be a useful production method in which the fast moving fragments are easy to identify and thus will enable a large number of experiments to study the properties of neutron-rich isotopes. Recent experiments using the VAMOS spectrometer to measure fission fragment yields from the reaction of $^{238}$U with $^{12}$C at near Coulomb barrier energies have demonstrated the advantage of inverse kinematics to study production mechanisms [2], and the properties of fission fragments [3]. However, in order to explore the properties of the most neutron-rich isotopes it is necessary to separate isotopes of interest from the more strongly produced nuclei.

---

[*] tarasov@nscl.msu.edu
[†] Present address : CEA DAM DIF, F-91297 Arpajon, France

In the present work a model [4] was developed to carry out fast calculations of the fusion–fission fragment cross sections with kinematics to facilitate studies of fusion-fission. The model was implemented in the LISE$^{++}$ package [5] and used existing analytical solutions for fusion–evaporation and fission fragment production mechanisms. In this work the advantages of in-flight fusion-fission in reverse kinematics to explore neutron-rich $55 < Z < 75$ region are shown by comparison to abrasion-fission and Coulomb fission. An important feature of reverse kinematics is that the excitation energy delivered to the heavy nucleus is relatively low even though the laboratory kinetic energy is high. The predictions are compared to the results of an experiment performed with the LISE3 separator [6] to separate and identify such fusion-fission products, generally verifies the new LISE$^{++}$ simulation with the fusion-fission model.

### 1.2. *Overall Reaction Scenario*

Fragment mass distributions from fusion-fission reactions have been extensively investigated [7,8] as they provide important information on the reaction dynamics and along with quasifission are the most important reaction channels involving the heaviest nuclei. In fact, the formation probability of super heavy elements in fusion reactions is completely determined by the fusion-fission process [9].

The present work describes a novel method to obtain additional information on the isotopic fission-fragment yields over the entire atomic-number range of the fission fragments (from $Z$=30 to $Z$=64), using inverse kinematics coupled with a fragment separator, in this case the LISE3 spectrometer.

In the present work the fusion-fission reactions were induced by a $^{238}$U beam at an energy of 24 MeV/u impinging on a 15 mg/cm$^2$ thick $^9$Be or natural $^{12}$C target. The main reaction channels and their general characteristics are given in Table 1 as a function of angular momentum. Note that the beam energy was approximately 20 MeV/u in the middle of the targets so that the excitation energies of the compound nuclei are moderate as seen in Table 1. The extreme asymmetry of these reactions hinders dissipative effects in the collision stage of the reaction, and strongly suppresses the quasifission mechanism [10,11].

Table 1. Main reaction channels and their characteristics for reactions of $^{238}$U (20 MeV/u) on Be and C targets as function of angular momentum calculated by the LISE$^{++}$ code for the fusion reaction mechanism [12]. The Sierk model [13] was used to estimate the fission barrier dependence on angular momentum.

|  | L (B$_{fis}$=0) | L critical | L direct | L max |
|---|---|---|---|---|
| Reaction characteristic | Fission barrier vanishes | Potential energy pocket vanishes | Corresponds to the interaction radius (max. s-wave barrier position) | Corresponds to the distance of minimum approach at grazing angle |
| Be-target | **67** | **75** | **78** | **89.2** |
| C-target | **63** | **87** | **99** | **117.1** |
| Reaction from previous L up to current L | Complete Fusion-Fission | Fast-Fission with high excitation (HE) sequential fission (FA) | Deep-Inelastic Collisions with HE sequential fission (DI) | Some part of Direct reactions go to sequential LE fission |
| Z of Fissile nucleus | Z of compound for targets Be: 96; C: 98 | Below projectile 85 < Z < 92 | | Around Z-projectile (92) |
| Fissile nucleus velocity | Compound velocity | Between compound and projectile velocities | | Close to projectile velocity |
| Excitation Energy of Fissile nucleus | at L=0 for C: 204.3 MeV Be: 166.6 MeV | Very broad energy range (30 MeV– Compound nucleus excitation energy) | | Low energy rang: 6-30 MeV |
| Z-distribution of fission fragments | One broad peak for Be: <Z>=48 C: <Z>=49 | Broad distribution with peak in Z~42-45 | | Two narrow peaks with Z around 38-40 and 52-54 |
| Reaction channel designation | FF (fusion-fission) | FA (fast-fission) | | IF (incomplete fusion) or DF |

## 2. Experiment

A $^{238}$U$^{58+}$ beam was accelerated to 24 MeV/u with an intensity on the order of 10$^9$ pps with the CSS1 and CSS2 cyclotrons at GANIL. The beam was directed onto the LISE3 target at an angle of 3° to prevent the unreacted beam from entering the spectrometer, as indicated in Figure 1.

Considering the high fission probability of the excited heavy nuclei produced in the collision, most of the reactions lead to the emission of fission fragments in a cone of about 10° around the beam direction. The small proportion that enter the LISE3 spectrometer were then identified by the combination of magnetic-rigidity $B\rho$, time-of-flight (*ToF*), total kinetic energy (*TKE*) and energy-loss ($\Delta E$) measurements. The identification of heavy ions using this technique is described in detail in the Appendix of Reference [14] . Two position-sensitive micro-channel plate detectors [15] measured the position of the particles $X_{31}$ and $X_{62}$ at the intermediate dispersive plane and the final focal plane, respectively, to deduce the magnetic rigidity of the particles.

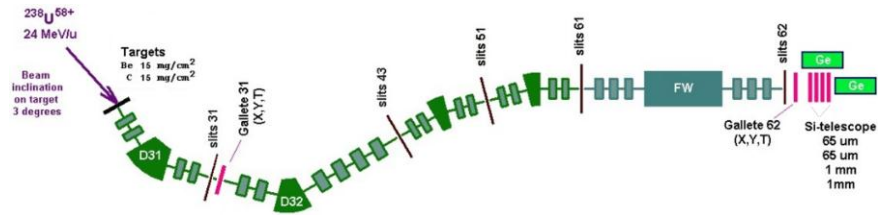

Fig. 1. Schematic layout of the LISE3 spectrometer with the detection equipment for the identification of fission fragments.

The spectrometer sections before and after the intermediate focal plane (slits 31) were set to the same magnetic rigidity. The position calibration of the micro-channel detectors was performed using slits placed in front of the detector. The spectrometer ion-optical parameters (dispersions and magnification) were calibrated using position measurements of different charge states of the primary beam. A stack of four silicon detectors was installed after the second microchannel plate detector (Gallete 62) to measure the energy loss and the energy of the ions. The time-of-flight of the fragments was measured between the micro-channel plate detector at the intermediate focal plane of the spectrometer (Gallete 31) and the first silicon detector at the final focal plane. The flight path was assumed to be independent of the measured position and equal to 32.423 m. The time-of-flight was calibrated by passing the uranium beam directly through the spectrometer.

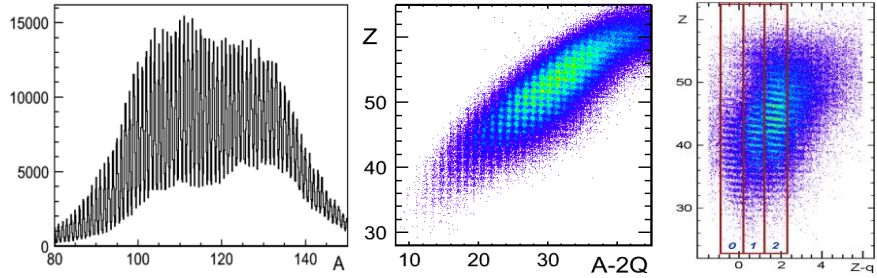

Fig. 2. Left: Mass distribution measured for the nominal magnetic rigidity $B\rho = 1.9$ Tm.: Z versus $A$-$2q$ (Middle) and Z versus $Z$-$q$ (Right) particle identification (PID) plots for the same spectrometer setting, see the text.

The observed mass distribution is displayed in panel (a) of Figure 2. A resolution of $\Delta A/A = 0.5\%$ FWHM was achieved, which provided good separation over the complete mass distribution. The atomic number Z of the fragments was identified with the energy-loss measurement in the first silicon detector of the silicon stack that had a thickness of 69 μm. The Z versus $A - 2q$ and Z versus $Z - q$ particle identification (PID) plots indicate the quality of the $A$, $Z$, and $q$ separation are shown in middle and right panels of Figure 2. The charge-state resolution obtained $\Delta q/q = 2\%$ was governed by the silicon-detector energy resolution. The isotopic identification was confirmed by the observation of the gamma-ray decay of isomeric states in several fission fragments with two germanium detectors placed around the silicon stack. Due to the limited beam-time the spectra had relatively low statistics, but as can be seen for the example in Figure 3, the decay of $^{128m}$Te, confirmed the particle identification scheme.

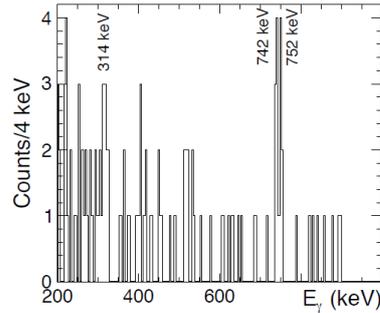

Fig. 3. Gamma-ray spectrum observed in coincidence with $^{128}$Te. The characteristic gamma lines of 314, 742 and 752 keV signal the decay of the known $T_{1/2} = 370$ ns state.

### 2.1. *Beam charge-state distribution*

The separator dispersion of 1.8 cm/% allowed the detection of several charge-states of the uranium beam within a magnetic-field setting and provided an absolute calibration of the spectrometer characteristics and a new measurement of the charge-state distribution. The magnetic rigidity of the spectrometer was scanned in order to cover the complete ionic charge-state distribution of the beam. The resulting charge-state distributions are displayed in Figure 4 and are compared to three different parameterizations typically used at this energy. Note that the beam enters the foils with $q = 58$ and the equilibrium charge state is near $q = 80$. Panels (a) and (c) show the beam charge-state distributions after passing through a thin Carbon or an Aluminized-Mylar layer, respectively, and it is clear that these thin layers barely strip the incoming beam These parameterizations assume that the material was thick enough to attain the equilibrium charge state and thus show a large discrepancy with the experimental. Panels (b) and (d), layers show the beam charge state distributions after passing through an aluminum or a beryllium layer, respectively, that is thick enough to reach the equilibrium charge-state, and the experimental data show better agreement with the data. In these cases the average charge-state increased from 58 to 76 and 79 after the Al and Be foils, respectively. The Leon parameterization [16] gives excellent results after the thick Al layer, whereas it is too high in the case of Be foil. In both cases, the Schiwietz model [17] gives a fair prediction of the average charge state, while the width of the distribution is too wide.

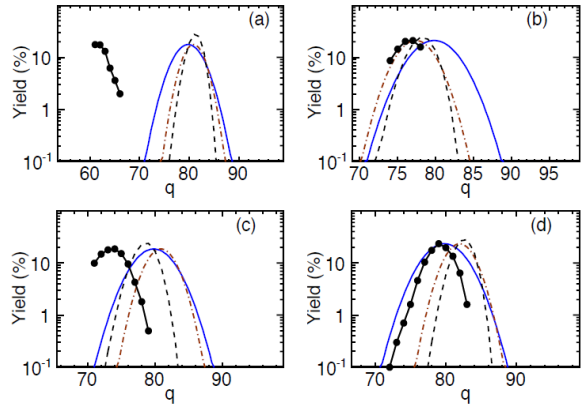

Fig. 4. Primary beam charge state distributions measured after passing through various materials. a) 40 μg/cm$^2$ C; b) 3 mg/cm$^2$ Al, c) 15 μg/cm$^2$ Mylar foil with 20 μg/cm$^2$ Al; d) 1.5 mg/cm$^2$ Be. The data are compared to parameter-zations of the charge-state distributions in the literature: Schiwietz [17] (solid blue line), Leon [16] (dotted-dashed red line), and Winger [18] (black dashed line).

## 3. Reconstruction of the fission fragment yields

The angular acceptance of the separator introduced cuts in the angular distribution of the fission fragments, as well as in their momentum distribution. The angle-aperture was ± 1°, and the momentum acceptance was set to ± 0.8%. The fragment production was measured at several values of magnetic rigidity, however due to the limited amount of beam-time, only four different rigidity values could be measured. In order to span as much as possible of the fragment momentum distribution, the four values were separated by approximately 5 % each.

The charge-state distributions of the fragments was estimated using the Schiwietz and Grande parameterization [17]. Figure 5 shows a comparison between the experimental charge-state distribution measured for a Zr and a Sn fragment from the ensemble of spectrometer settings along with the results of the simulation. The good agreement between the simulated and measured charge-state distributions gives confidence in the correct simulation of the kinematics and the charge state distribution.

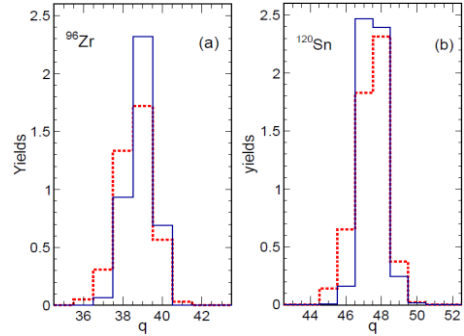

Fig. 5. Red dotted lines: charge state distributions of $^{96}$Zr and $^{120}$Sn with 15 mg/cm$^2$ Be-target in panels (a) and (b), respectively, measured for a magnetic rigidity of $B\rho = 1.9$ Tm. Dark blue solid lines are results of the simulation, see the text.

The yields measured at the four different spectrometer settings were normalized to the average incident beam intensity from measurements at the start and end of each run with a Faraday cup at the target position.

The values of the transmission calculated with LISE$^{++}$ for Abrasion-Fission and Complete Fusion–Fission reactions were used to deduce the fission production cross sections at each magnetic rigidity setting. Due to the fact the step size in magnetic rigidity between the measurements was the same, weighting the measured data by the transmission values is equivalent to integration of the data over rigidity. Abrasion-Fission reaction kinematics [19] was chosen to simulate the transmission of fast-fission (FA) products due to similarity of the characteristics of the products (see Table 1) as both sets of fissile nuclei are lighter than the projectile and have very broad excitation energies. It should be noted that in the LISE$^{++}$ fast analytical mode the reaction takes place in the middle of the target, therefore, in order to avoid too much averaging in transmission calculation by the rather thick targets (15 mg/cm$^2$), the targets were divided into 5 sections in the calculations. This method of combining transmissions for different reaction channels in order to obtain cross section is valid in the case with very similar contributions from both channels (here FF and FA). In the future it is necessary to evaluate the accuracy of this method.

## 4. Results and discussion

### 4.1. *Total cross sections*

The fission cross sections summed over all isotopes are found to be (3.6 ± 1.0) and (2.4 ± 0.7) barn for Be and C targets, respectively. Large systematic errors come from the beam monitoring that did not perform well for such relatively small primary beam intensities (< 20 enA). Total fission cross sections measured in this work at 20 MeV/u energy in the middle of the target far exceed the values of 2.00 ± 0.42 and 1.53 ± 0.15 barn obtained in high energy interactions of $^{238}$U (1 GeV/u) with deuterium [20] and hydrogen [21], respectively. Thus, there appears to be a significant fission contribution from the complete fusion channel at low energies that is absent in high energies.

### 4.2. *Elemental and neutron distributions*

The observed distributions as a function of atomic and neutron number of the fission fragments produced in this experiment are plotted in Figure 6, and the characteristics of the distributions are given in Table 2. These results are also compared to those obtained for the reactions $^{238}$U (1 GeV/u) on deutron [20] and proton [21] targets in the figure. The new results show that heavier fission fragments are produced at low energies and that this tendency is especially true in the case of beryllium target: on the average about 9 and 13 mass units fission fragment is heavier comparing with the proton [21] and deutron [20] targets, respectively. Also interesting, the shape of elemental distribution of fission fragments produced with the beryllium target is somewhat trapezoidal with a plateau from $Z = 46$ to 53 as compared to the more Gaussian shapes with the other targets.

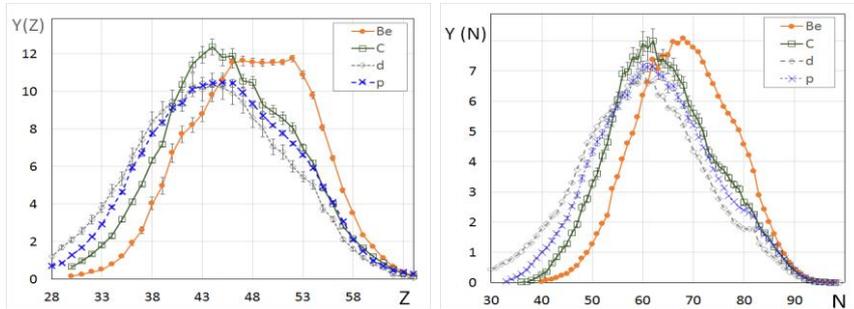

Fig. 6. Atomic (left) and neutron (right) numbers distributions of fission fragments produced by uranium at 24 MeV/u in this work with Be (solid dot) and C (open square) targets, along with those previously observed at 1 GeV/u with deuterium (open rhomb) [20] and hydrogen (cross) [21].

Table 2. Statistical characteristics of the atomic and neutron numbers distributions (see Figure 6) of fission fragments produced with uranium beams measured in the present work and compared with high energy results.

| Target | Energy | $<Z>$ | $\sigma_Z$ | $<N>$ | $\sigma_N$ | Ref |
|---|---|---|---|---|---|---|
| Be | 24 MeV/u | 48.01±0.22 | 6.03±0.17 | 68.29±0.18 | 9.30±0.14 | this work |
| C | 24 MeV/u | 45.75±0.21 | 6.40±0.16 | 64.16±0.17 | 10.22±0.13 | this work |
| p | 1 GeV/u | 44.93±0.20 | 7.00±0.15 | 62.60±0.16 | 11.18±0.12 | [21] |
| d | 1 GeV/u | 43.54±0.20 | 7.44±0.15 | 59.83±0.18 | 12.03±0.12 | [20] |

The widths of the fission fragment distributions produced by uranium in this work are shown as a function of atomic and neutron number and compared with high energy results on light targets [20,21] in Figure 7. The mean $N/Z$ ratios are shown in Figure 8 and clearly indicates that more neutron rich isotopes of elements below $Z = 48$ are produced with the beryllium target.

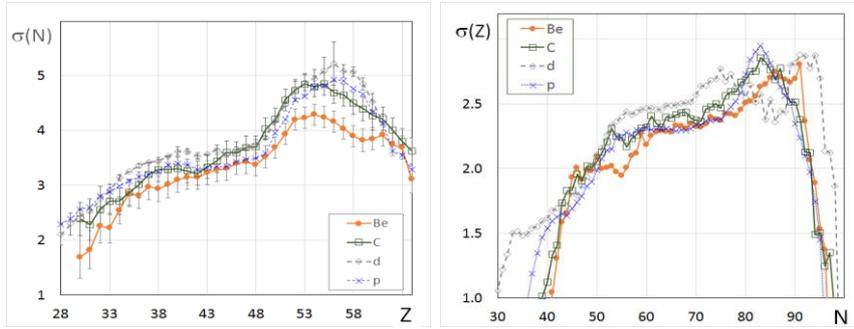

Fig. 7. Distributions of the neutron widths $\sigma_N$ (left) and atomic number widths $\sigma_Z$ (right) of fission fragments produced by uranium with energy 24 MeV/u in this work on Be (solid dot) and C (open square), and with energy 1 GeV/u with deuterium (open rhomb) [20] and hydrogen (cross) [21].

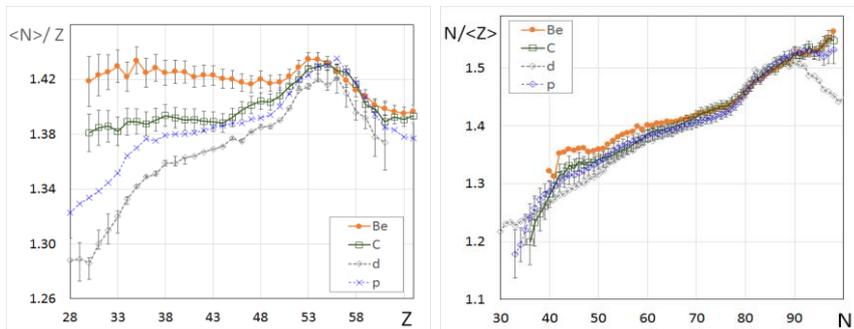

Fig. 8. Mean $N/Z$ ratio as a function of $Z$ (left) and $N$ (right) for fission fragments produced by uranium with energy 24 MeV/u in this work on Be (solid dot) and C (open square), and with energy 1 GeV/u with deuterium (open rhomb) [20] and hydrogen (cross) [21].

4.2.1. *Fission exit channels*

Given that the fission products are attributed to different reaction mechanisms with different fissile nuclei and excitation energies, it is interesting to see if the LISE$^{++}$ model can describe the observed distributions. The previous work [20,21] at high energy simply divided the distributions into a small component with asymmetric mass distribution from low energy fission and a large, single broad distribution created by a wide range of fission channels. The present low-energy data are more complex. In the first step of the analysis, only two high energy excitation fission channels were considered: (I) complete fusion forming the compound nucleus with a finite fission barrier (fusion-fission) and fission fragments centered at $Z = 48$ for the Be target and 49 for C target, plus (II) fusion at higher angular momenta leading to a nuclei without a fission barrier (fast-fission) and fission fragments centered at $Z = 44$ based on the high-energy analysis [20,21]. Normal distributions with an estimated width $\sigma_Z = 6$ were used with the constraint that the yield should not exceed experimental results. This constraint allowed one to find the positions of the low-energy asymmetric peaks in the rest of distributions after subtraction of Fusion-Fission and Fast-Fission components. The deduced Z-positions of the asymmetric fission (DF) at 40.0 and 53.5 (for carbon and beryllium targets, respectively) with $\sigma_Z = 2$ were used for the next step of minimization where the widths of the distributions for FA and DF positions with only steps 0.5. During the minimization in the case of Be-target all channels positions were conserved (FA at 44.0, DF at 40.0 and 53.5), whereas for Carbon target some shifts were obtained (FA at 43.5, DF at 41.5 and 53.0) with the reduced $\chi^2$ value of 4.3.

The contributions to the elemental distributions of the fission fragments from the different fission channels obtained from the analysis described above are shown in Figure 9. For the heavy asymmetric fragments produced at low excitation energy transfer (incomplete fusion) reactions, the maximum positions for both targets were found to be near $Z = 54$, which is agreement with the previous analysis by K.-H. Schmidt [22]. Note that the sum of low-energy peaks of 93.5 and 94.5 for Be and C targets, respectively, slightly exceed the number of protons in the projectile ($Z = 92$) which indicates a significant contribution from nucleon transfer (or incomplete fusion) from target to projectile in this energy domain.

As can be seen from Figure 9 the Fusion-Fission mechanism is responsible for high Z isotope production ($Z > 60$) for both targets, as discussed previously [4] and thus shows that in-flight fusion-fission is advantageous for exploring neutron-rich of high Z region due to heavier fissile nucleus.

Asymmetric fission at this energy, as well as at high energies, produced with light targets represents a small fraction of the total fission cross section. The main component in high energy experiments is sequential fission after abrasion of projectile (Abrasion-Fission), whereas in the current work fission after the complete fusion (FF) much dominates under other channels with the Be-target, and fusion-fast-fission dominates with the C target .

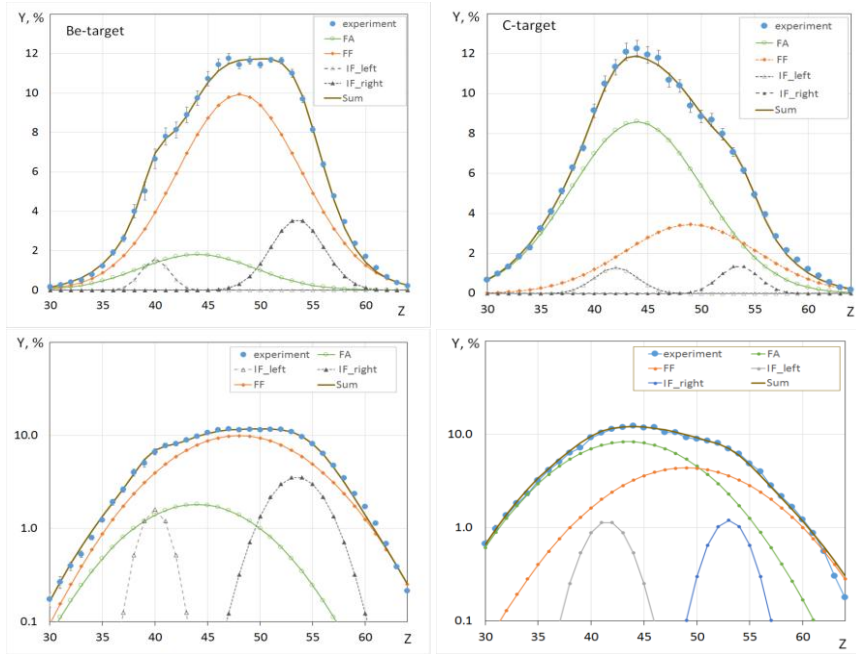

Fig. 9. Elemental fission yields (large blue solid circle) measured in the current work with beryllium (left plots) and carbon (right plots) targets. Top plots represent linear scale of vertical axes, bottom plots logarithmic scale correspondingly. Different fission channels contribution obtained from fitting are given in Table 3. Thick non-symbol solid lines are sum of fission channels. Fitting details are discussed in the text.

Table 3. Contribution of the independent fission channels determined in this work (see Figure 9), and corresponding them cross sections values obtained with use of the total measured fission cross sections.

| Fission Channel | Be target | | C target | |
|---|---|---|---|---|
| | Contribution | Cross section (mb) | Contribution | Cross section (mb) |
| Complete Fusion - Fission (FF) | 73.5 ± 2.2 % | 2680 ± 760 | 26.8 ± 2.6 % | 650 ± 200 |
| Fast-Fission (FA) | 12.5 ± 4.0 % | 460 ± 200 | 66.8 ± 5.5 % | 1620 ± 480 |
| Incomplete Fusion Fission (IF) | 13.9 ± 4.7 % | 500 ± 220 | 6.4 ± 4.8 % | 155 ± 100 |
| *Ratio FF/(FF+FA)* | *85.4 ± 5.1 %* | | *28.6 ± 3.3 %* | |

### 4.3. *Comparison with calculations*

The updated LISE++ code has been used to calculate reaction channels contribution in the current work and compared with experimental results including the partial wave description of the reaction mechanisms [12]. Recall first that there is no quasi-fission in such very asymmetric systems with light targets and that the fission channel completely dominates in de-excitation process. The partial cross sections at low angular momentum go into the complete fusion-fission channel (FF). The distribution of partial wave cross sections calculated by the LISE++ code for the reaction of $^{238}$U primary beam at energy 20 MeV/u with beryllium and carbon are shown in Figure 10, and their values are given in Table 4.

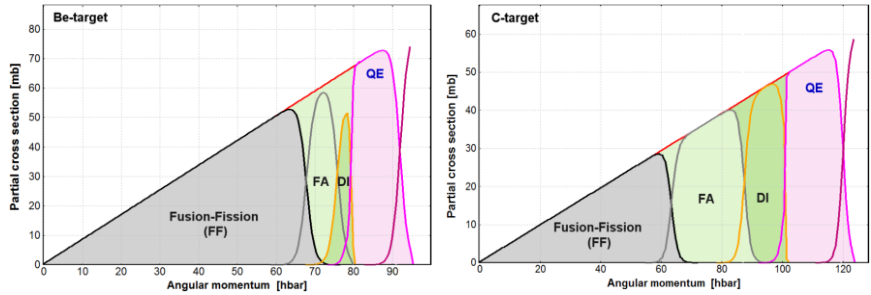

Fig. 10. Partial cross sections calculated by the LISE++ code for the reaction of $^{238}$U primary beam at energy 20 MeV/u with beryllium (left) and carbon (right) plots. Channels designation and angular momenta values are given in Table 1. Cross sections values are summarized in Table 4.

Table 4. Partial cross sections calculated by the LISE++ code for the reaction of $^{238}$U beam at energy 20 MeV/u with beryllium and carbon target (See Figure 10). The Sierk model [13] has been used to estimate a fission barrier dependence from angular momentum.

| Reaction channel | Be target | C target |
|---|---|---|
| Complete Fusion-Fission (FF) | 1987 | 1016 |
| Fast-Fission (FA+DI) | 643 (494+149) | 1455 (913+542) |
| Quasi-Elastic (QE) | 878 | 1013 |
| *Ratio FF/(FF+FA)* | *75.5%* | *41.1%* |

In general, the main trends of the experimental data (see Figure 9) are fairly well reproduced by the LISE++ calculations. The calculations show that (76%) fusion-fission should dominate fast-fission in the case of Be-target, whereas the picture changes rapidly if the target becomes a little bit heavier as it was in the carbon target case: domination of fast-fission (59%) under complete fusion-fission due to drastic increase of fast-fission contribution (see Figure 10). This analysis indicates that the difference in elemental experimental distributions of fragments produced with two different light targets could be explained by larger fast-fission component with C-target due to the formation of a significant number of nuclei with a vanishing fission barrier ..

## 5. Summary

Fusion-Fission reaction products produced by a $^{238}$U beam at 24 MeV/u on Be and C targets were measured in inverse kinematics with the LISE3 fragment separator. The identification of fragments was done using the d*E-TKE-Brho-ToF* method. Germanium gamma-detectors were placed in the focal plane near the Si stopping telescope to provide an independent verification of the isotope identification via isomer tagging. The experiments demonstrated excellent resolution in *Z*, *A*, and *q*. The results demonstrate that a fragment separator can be used to produce radioactive beams using fusion-fission reactions in inverse kinematics, and further that in-flight fusion-fission can become a useful production method to identify new neutron-rich isotopes, investigate their properties and study production mechanisms. Mass, atomic number and charge-state distributions are reported for both reactions.

The comparison of the experimental atomic and neutron number distributions combined with a partial-wave cross sections analysis indicate that the reaction mechanism changes substantially between the $^9$Be and the $^{12}$C targets, evolving from a complete fusion-fission to fast-fission.

The current analysis using two exit channels and a combined transmission shows only fair agreement between experimental data and calculations by LISE$^{++}$. The analysis was expanded to include a large contribution from fast-fission that occurs when the fission barrier vanishes dues to increasing angular momentum in the case of the carbon target.

The data suggest that complete fusion-fission is mostly responsible for production of fragments with *Z* > 60 in the case of target Be and C targets at 20 MeV/u  and this reaction may be used to produce neutron rich rare isotope beams in future studies.

**Acknowledgments**

The authors would like to acknowledge to Dr. G.N.Knyazheva (FLNR/Dubna) for fruitful discussions. This work was supported by the US National Science Foundation under Grants No. PHY-06-06007, No. PHY-10-68217, and No. PHY-11-02511.